\documentclass[amsmath, amssymb, preprintnumbers, showpacs, showkeys,aps,pra,superscriptaddress,twocolumn]{revtex4-1}

\usepackage{color} 
\usepackage{hyperref}
\usepackage{slashed}
\usepackage{graphicx}
\usepackage{amsmath}
\usepackage{latexsym}
\usepackage{epstopdf}
\usepackage{amsmath}
\usepackage{amssymb,amsmath}
\usepackage{mathtools}
\usepackage{bbm}
\usepackage{bm}
\usepackage{amsfonts}
\usepackage{amssymb}
\usepackage{calligra}
\usepackage{calrsfs}
\usepackage{makecell}
\usepackage{hyperref}

\usepackage[USenglish]{babel}

\newcommand{\beq}{\begin{equation}}
\newcommand{\eeq}{\end{equation}}

\newcommand{\be}{\begin{equation}}
\newcommand{\ee}{\end{equation}}

\newcommand{\bfr}{{\bf r}}

\newcommand{\tla}{\tilde{a}}
\newcommand{\tld}{\tilde{d}}

\newcommand{\tlr}{\tilde{r}}

\newcommand{\wbe}{\begin{widetext}}
\newcommand{\wee}{\end{widetext}}

\begin{document}
\title{Phases of  dipolar bosons in a bilayer geometry}

\author{Fabio Cinti}
\email{cinti@sun.ac.za}
\affiliation{National Institute for Theoretical Physics (NITheP), Stellenbosch 7600, South Africa}
\affiliation{Institute of Theoretical Physics, Stellenbosch University, Stellenbosch 7600, South Africa}

\author{Daw-Wei Wang}
\email{cdwwang@phys.nthu.edu.tw}
\affiliation{Physics Department, National Tsing-Hua University, Hsinchu, Taiwan}
\affiliation{Physics Division, National Center for Theoretical Sciences, Hsinchu, Taiwan}

\author{Massimo Boninsegni}
\email{m.boninsegni@ualberta.ca}
\affiliation{Department of Physics, University of Alberta, Edmonton, Alberta, Canada}



\begin{abstract}
We study by first principle computer simulations the low temperature phase diagram of bosonic dipolar gases in a bilayer geometry, as a function of  the two control parameters, i.e., the in-plane density and the interlayer distance.
We observe four distinct phases, namely paired and decoupled superfluids, as well as a crystal of dimers and one consisting of two aligned crystalline layers. 
A direct quantum phase transition from a dimer crystal to two independent superfluids is observed in a relatively wide range of parameters. No supersolid phase is predicted for this system.
\end{abstract}
\maketitle
\section {Introduction}
Quantum assemblies of particles featuring permanent electric or magnetic dipole moments are of interest for the intriguing, novel many-body
physical  effects that the anisotropic character of the interaction  may underlie \cite{baranov,lahaye}. In the simplest physical setting, a gas
of dipolar bosons is confined to two dimensions (2D), their dipoles all aligned perpendicularly to the plane by means of an external field; in this case, the interaction
between two particles is purely repulsive, decaying as $1/r^3$ at long distances -- neither short, nor quite long ranged.   Experimentally, a realization of such a system is possible with  molecules \cite{miranda}, ultra-cold Rydberg-excited atoms \cite{gallagher}, and ultra-cold bosonic gases of dysprosium \cite{pfauquasi2d} confined to 2D by means of an external harmonic trap.
The ground state ($T=0$) phase diagram of such a system  has been studied by Monte Carlo simulations \cite{buchler,Astrakharchik, mora,noi,jain}, yielding evidence of a first-order quantum phase transition between a superfluid and a crystal at  high density. 
\\ \indent
Of great interest is also the case of  a bilayer geometry, i.e., with dipolar particles (obeying either Fermi or Bose statistics) confined to two parallel planes.
In this case, if dipoles are aligned  as described above, the in-plane  interaction is purely repulsive, while that between particles in different planes is attractive at short distances.
The control parameters of this system, in the $T\to 0$ limit, are the in-plane density (or, equivalently, the mean interparticle distance $r_s$), assumed here to be the same for both planes, and the interlayer 
distance $d$. 
\begin{figure}[!t]
\includegraphics[width=8.3cm]{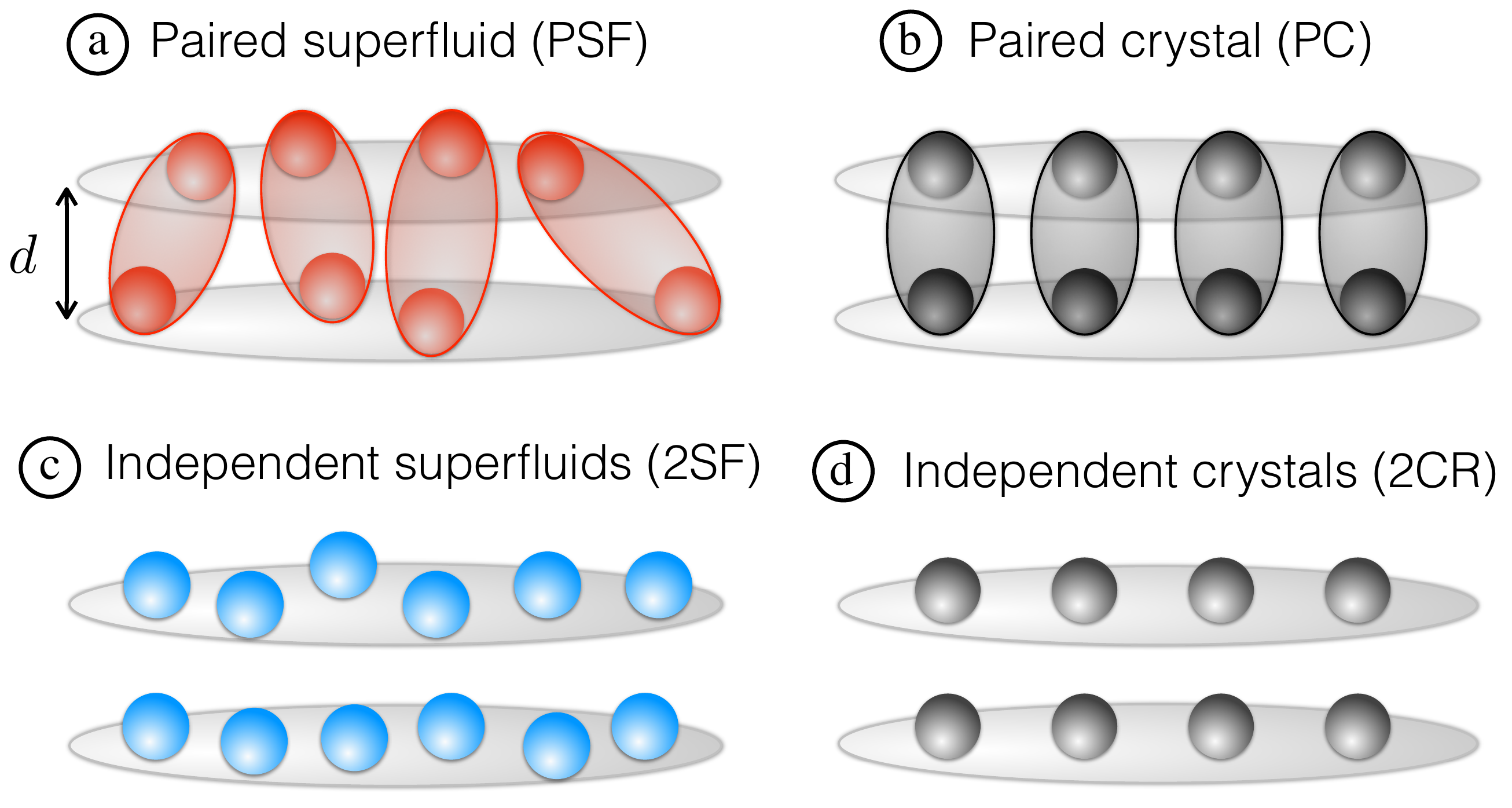} 
\caption{{\em Color online}. Phases of a bilayer bosonic dipolar system. There are two superfluid phases, one consisting of a single  superfluid of dimers (a), the other one comprising two independent 2D superfluids (b); (c)  shows a crystal of dimers, while (d) two aligned crystal layers.}
\label{four_phases}
\end{figure}
\\ \indent
The effect on the physics of the system of the  interaction between particles in different layers,  
depends on both $d$ and $r_s$ in a non-trivial way. In the two opposite limits $d >> r_s$ and $d << r_s$, one 
expects the physics to be the same as that of a single layer, in the first case because the two layers  decouple, in the second because the attraction between particles in different layers leads to the formation of  increasingly tightly bound pairs (dimers), acting like dipolar bosons of twice the mass  and
dipole moment of the original particles \cite{DW,DW2}.  On the other hand, in the intermediate regime in which $d\sim r_s$, one may expect novel phases to occur as a result of the competition between the repulsive in-plane and the (mostly) attractive out-of-plane interactions.
\\ \indent
In many respects, one can 
regard such a bilayer system as an ideal playground to gain general
understanding of the physics of composite particles (CPs), ubiquitous  in condensed matter  (one need only think of
Cooper pairs, polarons, excitons, composite fermions or Feshbach molecules) as well as in nuclear physics (hadrons) \cite{baryon}.
For example, it is clearly relevant to the physics of excitons, which are bosonic CPs expected to undergo 
Bose-Einstein condensation (BEC) at low temperature. In spite of tremendous experimental effort in the last decades \cite{exciton_BEC_Butov}, unambiguous observation of excitonic BEC is still elusive
\cite{exciton_BEC_comment,exciton_BEC_comment2}. It is also worthwhile mentioning that recent experimental advances in controlling 
ultra-cold dipolar atoms such as dysprosium \cite{pfauquasi2d} or erbium \cite{ferla} are paving the way to novel experiments 
on bi- or multi-layer geometries able to mimic the CPs physics here discussed. 
\\ \indent
The ground state phase diagram of dipolar bosons in a 2D bilayer geometry (in continuous space) has been studied 
by  Quantum Monte Carlo simulations \cite{Macia,filinov} at low in-plane density, where no crystallization occurs.
In this paper, we carry out a comprehensive study of the low temperature phase diagram of the system by means of Quantum Monte
Carlo simulations.  
\\ \indent
At low density, a $T=0$ quantum phase transition (previously observed by other authors \cite{Macia}) occurs  when the interlayer spacing $d$ is sufficiently small, compared to the interparticle distance $r_s$ (roughly $d/r_s \lesssim 0.5$); specifically, two decoupled 2D superfluids (hereinafter referred to as 2SF) transition into a phase featuring short-range pairing correlations between nearest neighboring particles in different layers. This phase, henceforth referred to as PSF (paired superfluid phase), has the character of a gas of tightly bound pairs (dimers) in the $d\to0$ limit; in the vicinity of the transition, on the other hand, pairing is more loosely defined,  and  the distinction between PSF and 2SF in the $T\to 0$ limit rests on the different superfluid properties (see below). 
\\ \indent
Increasing the density  while holding $d$ constant, has the effect of weakening  the effective interlayer interaction, as a result of which the 2SF phase  gains strength, extending to lower values of $d$ before transitioning into
a dimer crystal (PC).  Finally, in the high density limit superfluidity disappears, and the PSF is replaced at large $d$ by two independent crystals (2CR),  which are ``locked" into an aligned arrangement as a result of potential energy minimization, a fact already noticed by other authors \cite{pu} in the classical limit of Eq. (\ref{u}). 
All of these phases are schematically shown in Fig. \ref{four_phases}. At exactly $T=0$, the two crystalline phases are structurally indistinguishable. However, their melting behaviour at finite temperature is physically distinct, as will be illustrated below.
No supersolid phase is observed, consistently with the observation, repeatedly made in recent times, that a softening of the repulsive pairwise interaction is a necessary ingredient for the appearance of such a phase \cite{fabio,saccani,saccani2,jltp3,rmp,fabio2,jltp4}.
\\ \indent
The remainder of this paper is organized as follows: in Sec. \ref{method} we discuss the model and the methodology, with particular emphasis on the calculation
of the cogent quantities (mainly the superfluid density); in Sec. \ref{results} we illustrate our results. We outline our conclusions and 
discuss possible experimental observation of the phases described here in Sec. \ref{concl}.
\section{Model and Methodology}\label{method}
We consider an esemble of $2N$ Bose particles of spin zero,  mass $m$ and dipole moment $D$,  confined to either one of  two parallel planes at a 
distance $d$ from one another.
Each plane contains $N$ particles, a number that is fixed, i.e., there is no physical mechanism whereby particles can ``hop" from one plane to the other; 
all dipole moments are aligned in the direction perpendicular to the planes. 
\\ \indent
The  Hamiltonian of the system in dimensionless
units is the following:
\begin{eqnarray}\nonumber
\label{u}
\hat H=-\frac{1}{2}\sum_{i,\alpha}\nabla_{i,\alpha}^2 &+&\sum_{i\neq j,\alpha}\frac{1}{|\bfr_{i,\alpha}-\bfr_{j,\alpha}|^3} +\\
&+&\sum_{i,j}U_d({\bf r}_{i1},{\bf  r}_{j2})
\end{eqnarray}
with
\beq
U_d({\bf r},{\bf r}^\prime) = \frac{|\bfr-\bfr^\prime|^2-2d^2}{\left(|\bfr-\bfr^\prime|^2+d^2\right)^{5/2}}
\eeq
where $\bfr_{i,\alpha}$ is the position of the $i$th particle (dipole) of layer $\alpha=1, 2$. All lengths are expressed in terms of the characteristic length of the dipolar interaction, namely $a\equiv mD^2/\hbar^2$,
whereas $\epsilon\equiv (D^2/a^3)=\hbar^2/(ma^2)$ is the unit of energy and temperature (i.e., we set the Boltzmann constant $k_B=1$). The two control parameters  of the Hamiltonian (\ref{u}) in the $T\to 0$ limit, 
are the layer distance $d$ and the mean interparticle distance $r_s=(na^2)^{-1/2}$, where $n$ is the in-plane (2D) density \cite{notea}.
\\ \indent
The low temperature phase diagram of the system described by Eq.  (\ref{u}) has been studied in this work by means of  first principles numerical simulations, based on the continuous-space Worm Algorithm \cite{worm,worm2}.  Since this technique is by now fairly well-established, and extensively described in the literature, we shall not review it here. Details of the simulation are standard. In particular, we use a square cell with periodic boundary conditions in the two directions; the short imaginary time ($\tau$)  propagator utilized here is the usual one \cite{jltp}, accurate to order $\tau^4$;  all of the results presented here are extrapolated to the $\tau\to 0$ limit. Numerical results shown here pertain to simulations with a number of particles $N$ on each layer between 36 and 144.  
\\ \indent 
 Because we are mainly interested in the physics of the system in the $T\to 0$ limit,  we generally report here results corresponding to temperatures $T$ sufficiently low to regard them as essentially ground state estimates.  A quantitative criterion to assess whether the temperature $T$ of the simulation is sufficienty low, consists of monitoring the behaviour of specific physical quantities as a function of $T$. In the superfluid phase,  we consider ``ground state" estimates obtained at temperatures for which the computed superfluid fraction is within $\sim 5\%$ of its extrapolated $T=0$ value (as explained below, this depends on the phase which one is considering);  in the (non-superfluid) crystalline phase, the results that we furnish correspond to
a temperature $T \lesssim 10^{-2}\ \langle K\rangle $, $\langle K\rangle$ being the kinetic energy per particle. However, we also discuss the behaviour of the system as a function of temperature, notably the superfluid transitions and the melting of the crystal phases.\\ \indent
As stated above, the number of particles $N$ in each plane is constant, i.e., there is no physical mechanism allowing for interplane hopping; thus, the dipolar gases in the two planes are regarded as separate components, which requires the use of two separate ``worms'' \cite{worm2}, an especially important device 
in the 
study of paired superfluid phases.
\\ \indent
The use of a finite temperature technique to investigate what is essentially ground state physics might appear counterintuitive, considering that methods exist {\em in principle} purposefully designed to study the ground state of a many-body system (e.g.,  Diffusion Monte Carlo). In practice, however,  finite-temperature techniques 
typically prove superior in the investigation of Bose systems, even to determine ground state properties. This is mainly owing to the unbiasedness of finite temperature methods, which, unlike their $T$=0 counterparts, require no {\it a priori} physical input (e.g., a trial wave function), and are not affected by additional bias coming from, e.g., the finite size of the population of random walkers, like DMC \cite{bm,1dh2}. Moreover, finite temperature methods allow one to assess more easily and reliably quantities other than the energy, including off-diagonal correlations.
\\ \indent
As mentioned above, we compute the superfluid fraction of the system as a function of temperature, using the well-known ``winding number" estimator \cite{pollock}. In this case, it is necessary to distinguish between two types of superfluid phases (of the three that are known to occur in two-component Bose mixtures \cite{kuklov}), one in which superflow takes place independently in the two planes (top left in Fig. \ref{four_phases}), the other in which a superfluid of dimers occurs (top right in Fig. \ref{four_phases}) \cite{noteb}.  The two phases can be distinguished simply 
through the value of the in-plane superfluid fraction $\rho_S(T)$, which saturates to 100\% in the $T\to 0$ limit if two decoupled superfluids exist (one in each plane), but to 50\% in the presence of a superfluid of dimers, as a result of the twofold mass increase arising from the formation of the two-particle bound states.
\\ \indent
\section{results}\label{results}

\begin{figure}[!t]
\includegraphics[width=8.0cm]{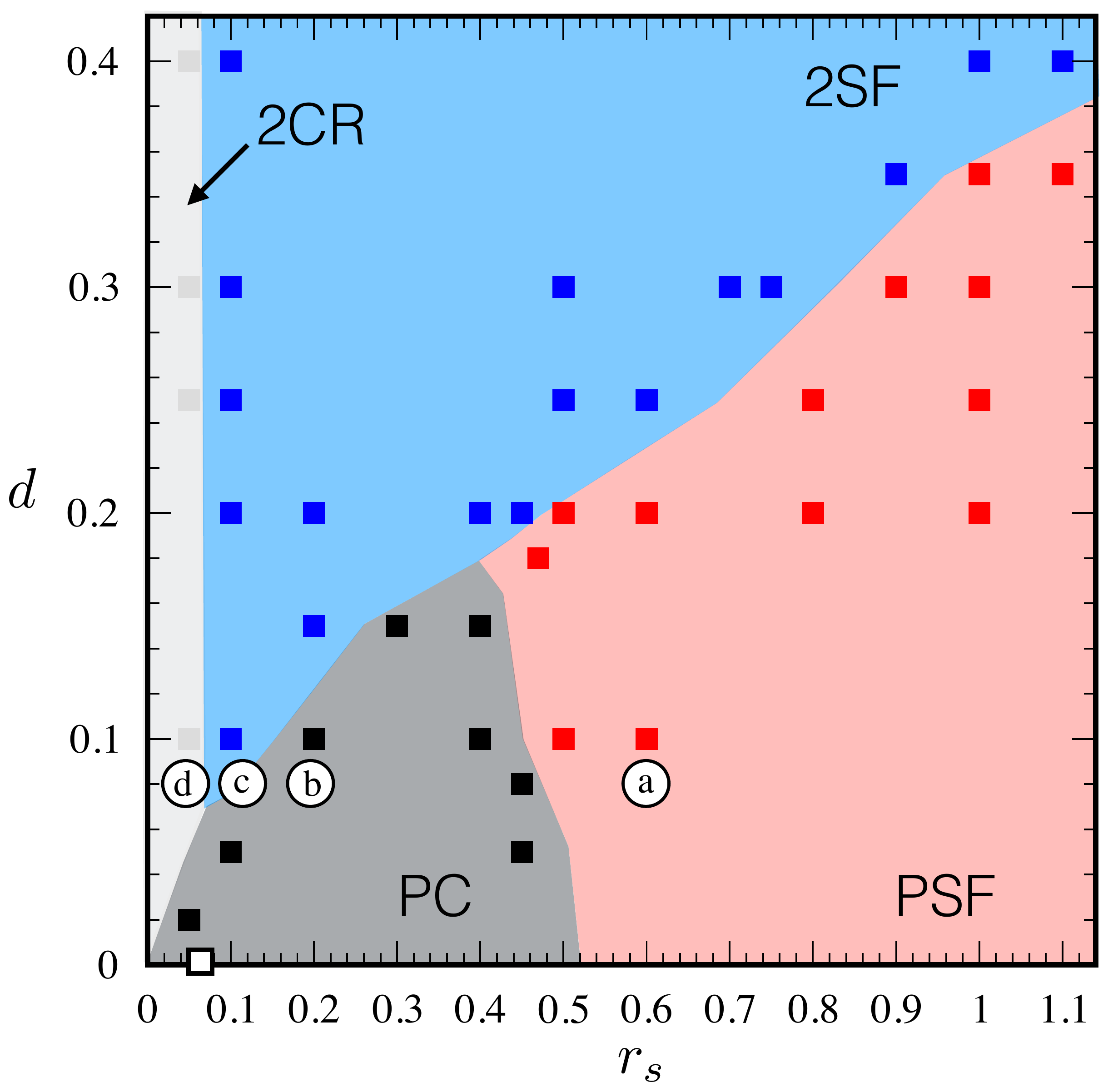}
\caption{{\em Color online}. Schematic ground state phase diagram of bosonic dipolar gases on a bilayer geometry, as a function of the interparticle distance 
$r_s$ and the interlayer separation $d$. Boxes refer to actual simulation results.
2SF stands for two decoupled 2D superfluids, PSF 2D  pair superfluid, PC for  pair crystal and 2CR for two separate 2D crystals. 
(a)-(d) are points for which pair correlation functions are shown in Fig. \ref{fig3}. 
Open square along the $d=0$ line represents the sigle layer crystal-superfluid phase transition as estimated in Ref.~\cite{noi}
}
\label{phase_diagram}
\end{figure}
Fig. \ref{phase_diagram} offers an overview of the phase diagram, with its four distinct phases, which we now discuss. We restricted our study to the $r_s \le 1$ region.
\subsection{Superfluid Phases}
\begin{figure}[!t]
\includegraphics[width=8.5cm]{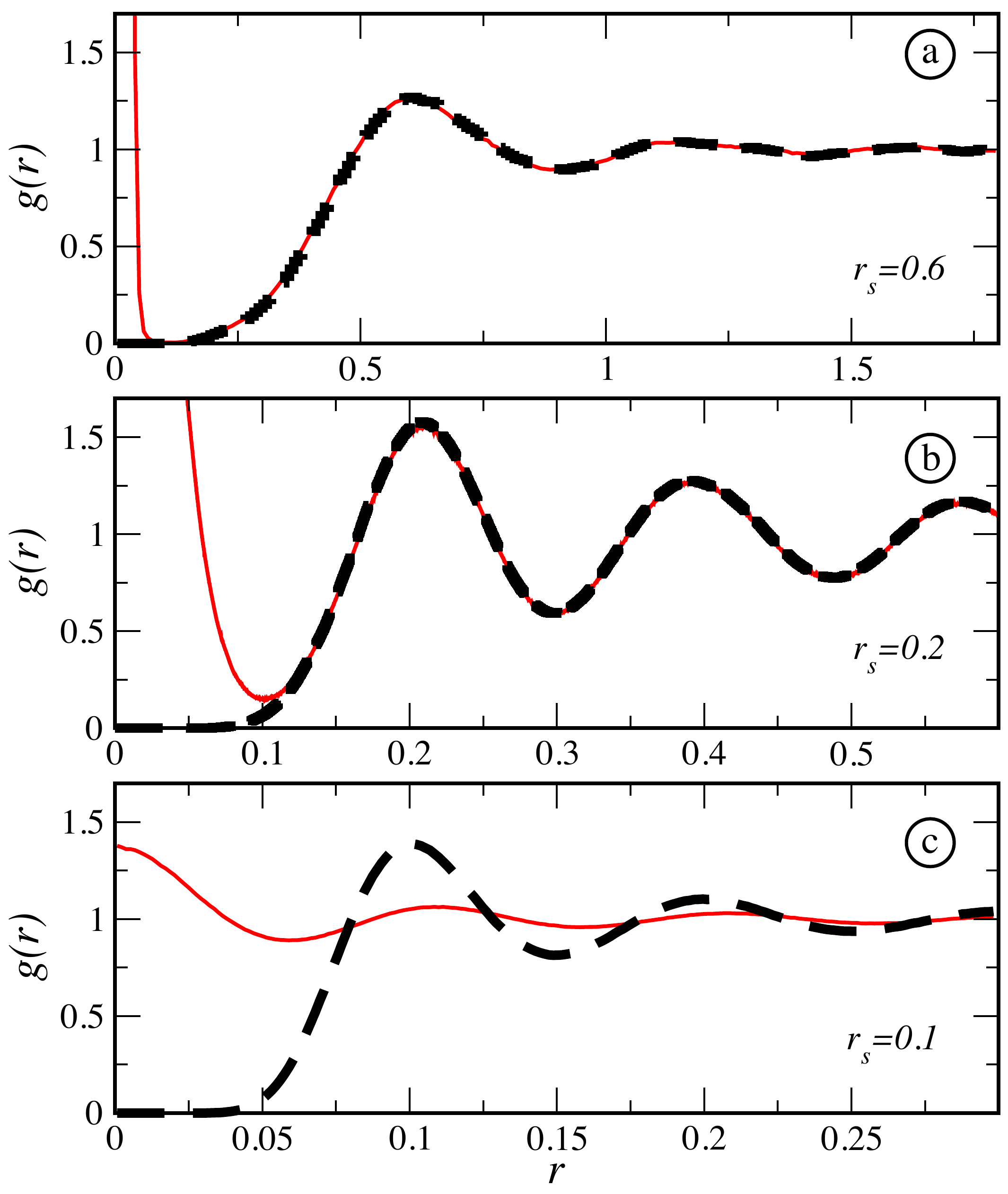}
\caption{{\em Color online.} Pair correlation functions $g_{\alpha\alpha}(r)$ (black dashed line) $g_{\alpha\beta}(r), \alpha\ne\beta$  (solid line), with $\alpha,\beta=1,2$ plane indices. 
The interlayer distance $d$=0.1.
Panel (a) refers to the pair superfluid phase (PSF) with $r_s=0.6$, (b) to the pair crystal (PC) with $r_s=0.2$ and (c) to two independent superfluids (2SF) with $r_s=0.1$. All of the curves shown here are representative of the $T=0$ limit. }
\label{fig3}
\end{figure}

At low density ($r_s \gtrsim 0.5$), the  system displays a 2D superfluid character, but the nature of the phase changes as the layers are brought sufficiently close. Specifically,  if the layers are far apart  ($d \gtrsim 0.4 \ r_s$), the system features two decoupled 2D superfluid gases (2SF in Fig. \ref{phase_diagram}). On the other hand, for close interlayer distances the short-range attractive well of the dipolar interaction between particles in different layers is deep enough that  bound states of particles in different layers form, and what one observes is a 2D superfluid phase of dimers (PSF in Fig. \ref{phase_diagram}).
\\ \indent
The two different regimes can be identified through the value of the in-plane superfluid density $\rho_S(T)$, which, as explained above, saturates to 100\% in the 2SF phase, but to 50\% in the PSF one, as a result of the doubling of the mass of each particle, which affects its diffusion in imaginary time. Their different structure can be assessed through the calculation of the pair correlation functions $g_{\alpha\beta}(r)$, where $\alpha,\beta=1,2$ are plane indices and where $r$ is a 2D distance.
\\ \indent
Fig. \ref{fig3} shows the $g_{\alpha\alpha}(r)$ and $g_{\alpha\beta}(r), \alpha\ne\beta$, pair correlation functions  pertaining to the thermodynamic points indicated as (a), (b) and (c) in Fig. \ref{phase_diagram}, i.e., $d=0.1$ in all cases. They are computed at sufficiently low temperature to be representative of the ground state of the system, i.e., the results do not change significantly, on the scale of the figures, if $T$ is further reduced.
\\ \indent
The $g_{\alpha\alpha}$ and $g_{\alpha\beta}$ shown in panel (a)  are essentially identical, except near the origin where $g_{\alpha\beta}$ (solid line) displays a strong peak, to signal the occurrence of a fluid phase of tightly bound pairs, with a short-range repulsion between pairs (as shown by both $g(r)$ going to zero at distance $\sim r_s/2$). These are the pair correlation functions characterizing the PSF phase.
Panel (c) of Fig. \ref{fig3}  shows instead the 2SF phase; here, the in-plane pair correlation function is that typical of a hard core fluid, with a broad main peak at $r=r_s$ followed by rapidly decaying oscillations at greater distances. Meanwhile, the corresponding function for particles in different planes only features a modest enhancement near the origin, on account of the attraction between particles in different layers when they are on top of one another, but otherwise is nearly constant, to indicate that the superfluid dipolar gases in the two layers are  decoupled.  
In other words, the main physical difference between 2SF and PSF is the existence in the latter of strong short-range correlations, which are always present in the PSF phase even when pairs are loosely bound. The character of the quantum phase transition between the PSF and the 2SF has been thoroughly discussed in Ref. \cite{barbara},  in which a study of a lattice version of Hamiltonian (\ref{u}) was carried out.
\\ \indent
The superfluid transition of both the 2SF and the PSF conforms to the 2D Berezinskii-Kosterlitz-Thouless (BKT) paradigm \cite{bere,kt}. While this is expected in the 2SF case, as it has already been verified for the single layer case \cite{fili2}, in the PSF regime the binding energy of a pair, of order $1/d^3$ in our units, is a few times the characteristic BKT superfluid transition temperature $T_{BKT}$, which is of the order of $1/r_s^2$. Thus, the system transitions to a normal fluid of pairs at finite $T$, dissociation occurring at higher $T$.
\\ \indent
It is worth noting that the physical behaviour of the system in the superfluid part of the phase diagram is {\it not} independent of $r_s$. Specifically, while in the regime considered here (i.e., $r_s < 1$)  the physics of  is that of a 2D quasi-BEC of (relatively) tightly bound pairs,
in the $r_s >> 1$ limit (not investigated here)  the binding energy of the dimers  decreases exponentially \cite{Simon,Pikovski}  with the interlayer distance $d$; thus,  the spatial size of the dimer wave function can become comparable to the interparticle distance, and the physics of the system approaches that of a BCS superconductor.
\\ \indent
In the vicinity of the 2SF/PSF quantum phase transition, the peak at the origin of the $g_{\alpha\beta}$ correlation function in the PSF phase tends to get weaker in the $T\to 0$ limit, due to both quantum exchanges and zero-point motion; however, within the range of density considered in this work, in no case do we observe thermal reentrance of the PSF phase (described, e.g., in Ref. \cite{kuklov}). 
However,  the method utilized in this work does not allow us to exclude a fundamental change of the character of the phase at temperatures unattainable in practice, given the current computational resources. In any case, we note that this result seems consistent with the findings of Ref. \cite{barbara}. Indeed, the main difference between our results and theirs is that no supersolid phase is observed in the continuum, reflecting an important, intrinsic difference between lattice and continuum Hamiltonians \cite{rmp}. Indeed, the lattice version of  (\ref{u}) features a supersolid phase even on a single layer \cite{picon}.
\\ \indent
\subsection{Crystal Phases}
\begin{figure}[!t]
\includegraphics[width=8.4cm]{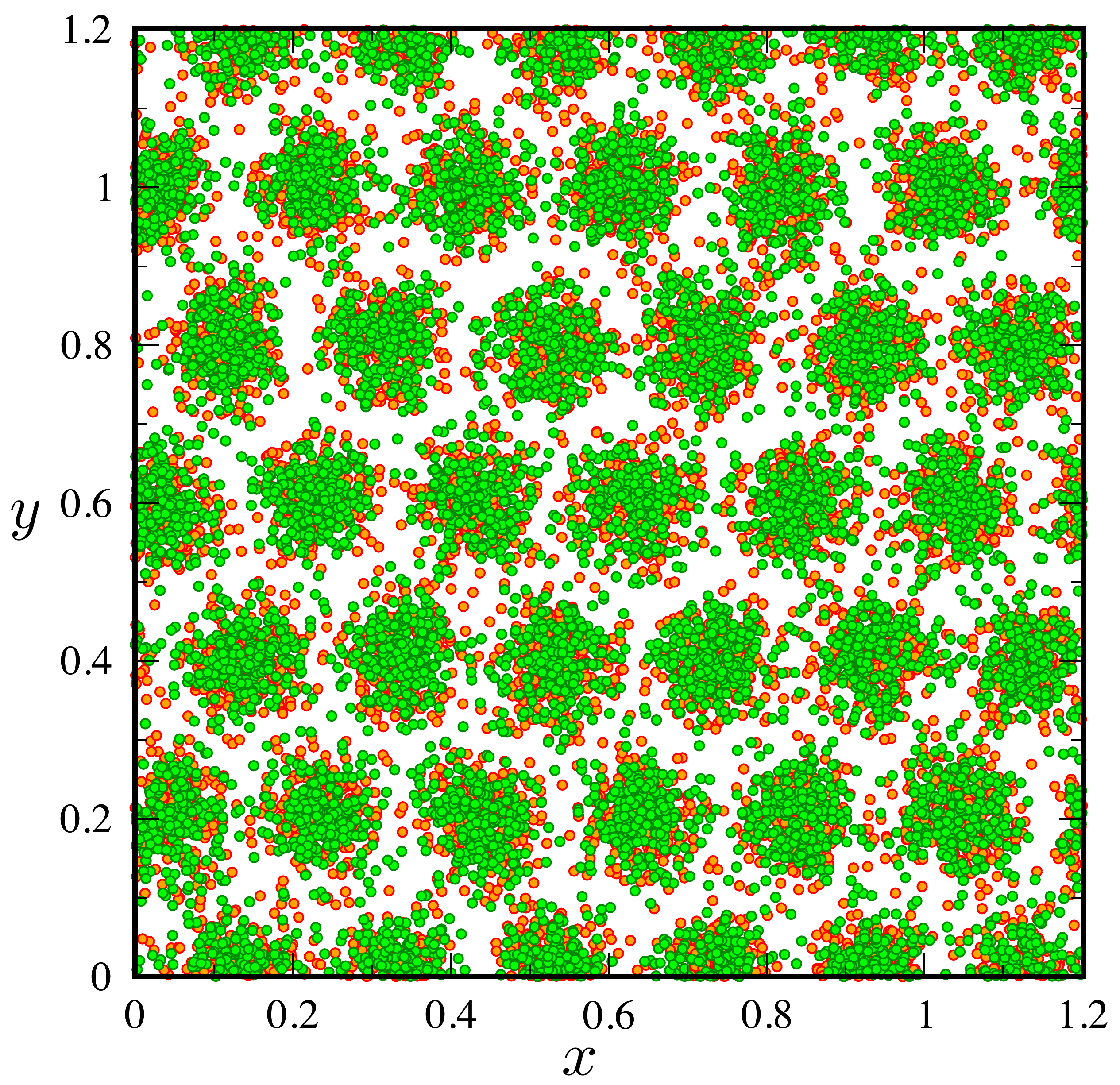} 
\caption{{\em Color online.} Snapshot of many-particle configurations (world lines) for a bilayer system with
mean interparticle distance $r_s=0.2$ and interlayer separation $d=0.1$, at temperature $T=0.1$ in the units adopted here (see text). Different colors refer to particles in different layers.}
\label{fig4}
\end{figure}

For $r_s \lesssim 0.5$ and sufficiently low $d$ ($d\lesssim 0.5\ r_s$), a crystalline phase of dimers arise (PC in Fig. \ref{phase_diagram}). The pair correlation functions for this phase are shown in Fig. \ref{fig3} (b). There is a feature in common with the  PSF phase, namely the strong peak at the origin (solid line) and the fact that the two $g(r)$ are on top of one another for distances greater than $\sim 0.5\ r_s$.
Unlike those of the PSF phase, however, the pair correlation functions for this case display the persistent, marked oscillations that are typical of the crystalline phase. Just like for the PSF phase, the strong peak at the origin indicates the formation of  tightly bound pairs. Crystallization can be rather easily detected by visual inspection of  the many-particle configurations generated in the course of a simulation; a typical example is shown in Fig. \ref{fig4}, displaying an instantaneous snapshot of particle world lines. It is important to note that the crystalline arrangement shown in Fig. \ref{fig4} occurs {\em spontaneously}, i.e., it is not initially imposed at the beginning of the simulation; the fact that particles in the two layer align nearly perfectly makes only one of the two layer clearly visible.
The factor 8 difference between the freezing density of the single-layer  system and that of the dimer one in the $d\to 0$ limit is a consequence of the doubled particle mass and fourfold increase of the strength of the dipolar coupling, as each dipole is doubled.
\\ \indent
As $r_s \to 0$, two main physical effects occur, i.e., {\it a}) the in-plane repulsive interaction increases, driving the system toward crystallization in each plane, and {\it b}) the interlayer interaction is weakened. The weakening takes place as particles in one layer increasingly feel the effect of the nearest neighboring particles in the other layer,  as opposed to only (or, mainly) the one directly above, as is the case at low density. The overall result is that of a softening of the interlayer interaction, if the distance $d$ is kept constant.
The most important physical consequence is a strengthening of the 2SF phase, which progressively extends its domain of stability at low temperature to lower 
values of $d$, until $r_s$ reaches a value for which in-plane freezing into a triangular lattice begins to occur (as shown in Fig. \ref{phase_diagram}). This value has been recently accurately estimated to be close to $r_s=0.064$ \cite{noi}.
Our simulation confirms in-plane crystallization for this value of $r_s$, essentially independently of $d$, for $d\gtrsim 0.1$. The crystals in the two layers are ``locked" into a configuration in which each particle in one layer sits above one in the other layer, as this minimizes the potential energy. We refer to this crystalline phase, which is physically distinct from the PC one, as consisting of two independent 2D crystals (2CR).\\ \indent
In the $T\to 0$ limit, the 2CR and PC phases are structurally indistinguishable. Neither the pair correlation functions nor snapshots like that shown in Fig. \ref{fig4}  show any qualitative or quantitative differences. Rather, as observed also in Ref. \cite{pu}, it the behaviour of the system at finite $T$ that allows one to draw a physically meaningful distinction between the two crystalline phases. 
\begin{figure}[!t]
\includegraphics[width=8.4cm]{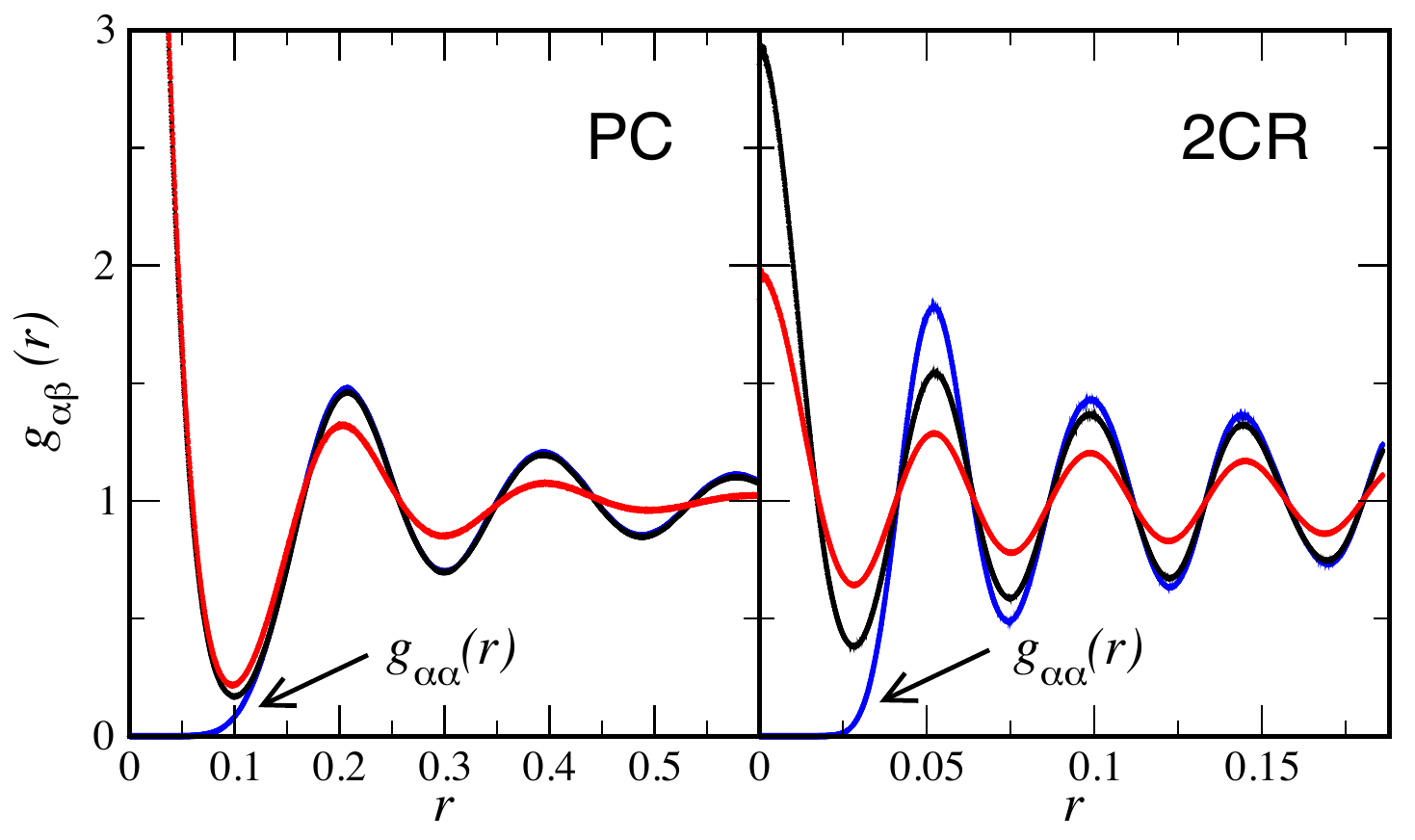} 
\caption{{\em Color online.} Pair correlation function $g_{\alpha\beta}(r)$, $\alpha\ne\beta$ at different temperatures, for $d=0.1$. {\em Left}:  $r_s=0.2$, and $T$=1, 100 in the units utilized in this work. {\em Right}: $r_s=0.05$, $T=40, 125$. Lower peaks correspond to higher $T$.  Also shown is the corresponding $g_{\alpha\alpha}(r)$ at the lowest temperature for each case (arrows). In the results shown  left panel, the $g_{\alpha\alpha}$  cannot be distinguished from the
$g_{\alpha\beta}$ at the same temperature, for $r \gtrsim 0.1$.
  }
\label{fig5}
\end{figure}
\\ \indent
This is illustrated in Fig. \ref{fig5}, which shows the temperature behaviour of the pair correlation function $g_{\alpha\beta}$, $\alpha\ne\beta$, for two different cases, corresponding to PC (left) and 2CR (right) ground states. Left panel shows $g_{\alpha\beta}$ for $d=0.1$ and $r_s=0.2$, at the two temperatures $T$=1, 100 in the units utilized here;  right panel shows results for $d=0.1$, $r_s=0.05$ and $T$=40, 125.  In the results shown in the left panel, the $g_{\alpha\alpha}$ and $g_{\alpha\beta}$ at the two temperatures are virtually indistinguishable, for $r\gtrsim r_s/2=0.1$; at shorter distance, the $g_{\alpha\beta}$ develops a peak, as particles in different layers line up, whereas the $g_{\alpha\alpha}$ vanishes as a result of the in-plane, hard core repulsion of the dipolar interaction. As the temperature is raised, both the $g_{\alpha\alpha}$ and $g_{\alpha\beta}$ lose structure, as the crystal order characterizing the ground state progressively disappears, but they remain indistinguishable above $r_s\sim r_s/2$. This is consistent with the melting of  a system of tightly bound dimers.
\\ \indent
Let us now examine the very different behaviour shown in the right panel. Here too, as expected, in the $T\to 0$ limit, $g_{\alpha\beta}$ and $g_{\alpha\alpha}$ become identical for
$r\gtrsim 0.025=r_s/2$, and display the same features as the curves in the left panel  for $r \lesssim r_s/2$; however, as the temperature is raised, the $g_{\alpha\beta}$ quickly loses structure, while the $g_{\alpha\alpha}$ changes very little in the temperature range shown. Visual inspection also confirms that crystalline order persists in both planes. However, at finite temperature the simulated, finite-size crystals in the two layers can shift with respect to one another, as a result of the weak interlayer potential energy of attraction, which causes the $g_{\alpha\beta}$ to be almost ``flat" (i.e., nearly constant at a value unity) at a temperature $T$=250. All of this shows that the physics of the system is essentially that of two independent layers. The melting of the in-plane crystal takes place at considerably higher temperature than those shown in the right panel of Fig. \ref{fig4}.
\\ \indent
Returning to the ground state phase diagram of Fig. \ref{phase_diagram}, an interesting feature arising from the competing effects of in-plane repulsion and out-of-plane attraction, 
is the presence of a region, e.g, $0.05\le r_s\le 0.2$, $d=0.1$, inside which,  on increasing the density (i.e., $r_s\to 0$), first the dimer crystal quantum melts into two decoupled superfluids, which then successively crystallize again into the 2CR phase. At $T=0$, the PC phase can quantum melt into a PSF or 2SF, as the layers are moved away from one another, or transitions into 2CR. It is worth mentioning that the stability of the 2SF phase, near the 2CR and PC phase boundaries (e.g., $r_s=0.1,\, d=0.1$, see Fig. \ref{phase_diagram}), is crucially underlain by quantum-mechanical exchanges; indeed, simulations treating particles as {\em distinguishable} yield a stable PC phase in a considerably more extended region of the phase diagram, as already remarked in previous works, for dipolar Bose  systems \cite{jam}.
\\ \indent
We conclude this section by discussing the melting of the PC and 2CR phases. Our simulations show that the PC phase always melts into a dimer fluid, either normal or superfluid depending on the density. In particular, in the $d\to 0$ limit, when the dimers are strongly bound, the physics of the system reproduces that of a single layer, which is of course also approached in the $d >> r_s$ limit, the only difference between the two regimes being a rescaling of the unit of length by a factor 8, as explained above.
We obtained in this work numerical evidence  of melting of the single-layer system into a superfluid, close to the $T=0$ melting density (i.e., $r_s\sim 0.5$). On the other hand, the 2CR phase is always found to melt into two independent normal fluid phases on the two layers.

\section{Conclusions}\label{concl}
In this work, we have employed exact numerical methods to investigate the low temperature phase diagram of dipolar bosons in a bilayer geometry, all dipoles aligned perpendicularly to the planes.
In the two opposite limits in which the in-plane mean interparticle distance $r_s$ is either much less or much greater than the interlayer separation $d$, the
physics of the system is that of a single-layer system \cite{buchler,noi}. 
On the other hand, as a result of the competition between the in-plane (repulsive) and the out-of-plane (attractive) interactions, 
the intermediate regime ($d\approx r_s$) gives rise to considerably more complex and interesting  physics. 
In particular, at low density ($r_s\gtrsim 0.5$) a quantum phase transition occurs as $d \lesssim r_s/2$ from a phase consisting of two independent 
superfluid 2D gases, to a superfluid phase of bound pairs of particle in different layers.
No reentrant behaviour of either phase was observed at finite $T$.
\\ \indent
At higher density, we observe two distinct solid phases, physically related to the superfluid ones, namely one of tightly bound dimers, which arises when the interlayer separation is less than the mean interparticle distance, and one comprising two independent 2D crystalline layers. An interesting feature of the phase diagram is the direct transition of the system from a crystalline phase of dimers into one of independent 2D superfluids, which is observed in a rather wide range of parameters.
\\ \indent
In terms of possible experimental realization,
one can estimate characteristic physical values for the parameters of  Eq. (\ref{u}), e.g., by considering a realistic polar molecule, say SrO, for which $a\sim 122.2$ $\mu$m for a fully polarized state with $D=8.9$ Debye. Considering that both  the interlayer and mean interparticle distances are of the order of a fraction of a $\mu$m in typical experiments, one can easily imagine tuning $r_s$ and $d$ in a rather wide range.
\section*{Acknowledgments}
We acknowledge fruitful discussion with F. Mezzacapo, J.-S. You, and Y.-Y. Tian. This work was supported in part by the Natural Science and Engineering Research Council of Canada. Computing support of Westgrid is gratefully acknowledged. DDW is supported by research grant of NCTS and MoST in Taiwan.
 


\begin{thebibliography}{99}

\bibitem{baranov}
M. A. Baranov, Phys. Rep. {\bf 464}, 71 (2008). 

\bibitem{lahaye}
T.  Lahaye, C. Menotti, L.  Santos, M. Lewenstein, and T. Pfau, Rep. Prog. Phys. {\bf 72}, 126401 (2009).

\bibitem{miranda}
M. H. G. de Miranda,	 A. Chotia, B. Neyenhuis,	D. Wang,	G. Qu\'em\'ener, S. Ospelkaus,	J. L. Bohn, J. Ye, and D. S. Jin, Nature Physics {\bf 7}, 502?507 (2011).

\bibitem{gallagher}
See, for instance, T. G. Gallagher, {\em Rydberg Atoms} (Cambridge University Press, Cambridge, 1994).

\bibitem{pfauquasi2d}
H. Kadau,	M. Schmitt, M. Wenzel, C. Wink, T. Maier,	I. Ferrier-Barbut, and T. Pfau,  Nature {\bf 530}, 194 (2016).

\bibitem{buchler}
H.-P. B\"uchler, E. Demler, M. Lukin, A. Micheli, N. Prokofev, G. Pupillo, and P. Zoller, Phys. Rev. Lett. {\bf 98}, 060404 (2007).

\bibitem{Astrakharchik}
G. E. Astrakharchik, J. Boronat, I. L. Kurbakov, and Yu. E. Lozovik, Phys. Rev. Lett. {\bf 98}, 060405 (2007).

\bibitem{mora}
C. Mora, O. Parcollet, and X. Waintal, Phys. Rev. B {\bf 76}, 064511 (2007).

\bibitem{noi}
S. Moroni and M. Boninsegni, Phys. Rev. Lett. {\bf 113}, 240407 (2015).

\bibitem{jain}
P. Jain, F. Cinti, and M. Boninsegni, Phys. Rev. B {\bf 84}, 014534 (2011).

\bibitem{DW}
D.-W. Wang, M. D. Lukin, and E. Demler, Phys. Rev. Lett. {\bf 97} 180413 (2006).

\bibitem{DW2}
D.-W. Wang, Phys. Rev. Lett. {\bf 98}, 060403 (2007).

\bibitem{baryon}
See for example, B. R. Martin, {\it Nuclear and Particle Physics}, John Wiley \& Sons (2006).

\bibitem{exciton_BEC_Butov}
L. V. Butov, C. W. Lai, A. L. Ivanov, A. C. Gossard, and D. S. Chemla, Nature {\bf 417}, 47 (2002); A. A. High, J. R. Leonard, M. Remeika, L. V. Butov, M. Hanson, and A. C. Gossard, “Condensation of Excitons in a Trap,” Nano Lett. {\bf 12}, 2605 (2012).

\bibitem{exciton_BEC_comment}
See for example, P. B. Littlewood and P Eastham, NATO Science Series {\bf 81}, 133 (2000);
D. Semkat, S. Sobkowiak, G. Manzke, and H. Stolz, Nano Lett., {\bf 12}, 5055 (2012); M. S. Fuhrer, and A. R. Hamilton, Physics {\bf 9}, 80 (2016).

\bibitem{exciton_BEC_comment2}
Recent progress for quantum Hall excitons may have some new development, but full many-body theory in the high density regime seems still unclear. See J. P. Eisenstein and A. H. MacDonald, Nature {\bf 432}, 691 (2004); P. Bhattacharya, {\it et. al.} Phys. Rev. Lett. {\bf 110}, 206403 (2013), and C. Schneider, {\it et. al.}, Nature {\bf 497}, 348 (2013), 

\bibitem{ferla} 
L. Chomaz, S. Baier, D. Petter, M.J. Mark, F. W\"achtler, L. Santos, and F. Ferlaino, Phys. Rev. X {\bf 6}, 041039 (2016). 


\bibitem{Macia}
A. Macia, G. E. Astrakharchik, F. Mazzanti, S. Giorgini, and J. Boronat, Phys. Rev. A {\bf 90}, 043623 (2014).

\bibitem{filinov}
A. Filinov, Phys. Rev. A {\bf 94}, 013603 (2016).

\bibitem{pu}
X. Lu, C.-Q. Wu, A. Micheli, and G.  Pupillo, Phys. Rev. B {\bf 78}, 024108 (2008).

\bibitem{fabio}
F. Cinti, P. Jain, M. Boninsegni, G. Pupillo, A. Micheli, and P. Zoller, Phys. Rev. Lett. {\bf 105}, 135301 (2010).

\bibitem{saccani}
S. Saccani, S. Moroni and M. Boninsegni, Phys. Rev. B {\bf 83)}, 092506 (2011).

\bibitem{saccani2}
S. Saccani, S. Moroni and M. Boninsegni,  
Phys. Rev. Lett. {\bf 108}, 175301 (2012). 

\bibitem{jltp3}
M. Boninsegni, J. Low Temp. Phys. {\bf 168}, 137 (2012).

\bibitem{rmp}
M. Boninsegni and N. Prokof'ev, Rev. Mod. Phys. {\bf 84}, 759 (2012). 

\bibitem{fabio2} 
F. Cinti, T. Macr\`i, W. Lechner, G. Pupillo, and T. Pohl; Nat. Commun. {\bf 5}, (2014).

\bibitem{jltp4}
M. Boninsegni, J. Low. Temp. Phys.  {\bf 184}, 1071 (2016).

\bibitem{notea}
We are obviously implicitly assuming that the layer width is much smaller than both $r_s$ and $d$, i.e., the layers can be regarded as very nearly 2D.

\bibitem{worm} M. Boninsegni, N. Prokof’ev, and B. Svistunov, Phys.
Rev. Lett. {\bf 96}, 070601 (2006).

\bibitem{worm2} M. Boninsegni, N. V. Prokof’ev, and B. V. Svistunov,
Phys. Rev. E {\bf 74}, 036701 (2006).

\bibitem{jltp}
See, for instance, M. Boninsegni, J. Low Temp. Phys.  {\bf 141}, 27 (2005).

\bibitem{bm}
M. Boninsegni and S. Moroni, Phys. Rev. E {\bf 86}, 056712 (2012).

\bibitem{1dh2}
M. Boninsegni, Phys. Rev. Lett. {\bf 111}, 235303 (2013).

\bibitem{pollock}
E. L. Pollock and D. M. Ceperley, Phys. Rev. B {\bf 36}, 8343 (1987).

\bibitem{kuklov}
A. Kuklov, N. Prokof'ev and B. Svistunov, Phys. Rev. Lett.  {\bf 92}, 030403 (2004).

\bibitem{noteb}
No phases featuring ``super-counterflow'' is observed here. 

\bibitem{barbara}
A. Safavi-Naini, S. G. S\"oyler, G. Pupillo, H. R. Sadeghpour
and B. Capogrosso-Sansone, New J. Phys. {\bf 15}, 013036 (2013).


\bibitem {bere} V. L. Berezinskii, Sov. Phys.  JETP {\bf 32}, 493 (1971).

\bibitem{kt} J. Kosterlitz and D. Thouless, J. Phys. C {\bf 6}, 1181 (1973).

\bibitem{fili2}
A. Filinov, N. V. Prokof’ev  and M. Bonitz, Phys. Rev. Lett. {\bf 105}, 070401 (2010).

\bibitem{picon}
L. Pollet, J. D. Picon, H.-P. B\"uchler and M. Troyer, Phys. Rev. Lett. {\bf 104}, 125302).
\bibitem{Simon}
B. Simon, Annals of Physics {\bf 97}, 279 (1976).

\bibitem{Pikovski}
A. Pikovski, M. Klawunn, G. V. Shlyapnikov, L. Santos, Phys. Rev. Lett. {\bf 105} 215302 (2010).

\bibitem{jam}
M. Boninsegni, L. Pollet, N. Prokof'ev and B. Svistunov, Phys. Rev. Lett. {\bf 109}, 025302  (2012).

\end{thebibliography}
\end{document}